# HEP computing collaborations for the challenges of the next decade

Contacts: Simone Campana (Simone.Campana@cern.ch), Zach Marshall (ZLMarshall@lbl.gov), Alessandro Di Girolamo (Alessandro.Di.Girolamo@cern.ch), Heidi Schellman (Heidi.schellman@oregonstate.edu), Paul Laycock (laycock@bnl.gov), Graeme A Stewart (graeme.andrew.stewart@cern.ch).

## Abstract

Large High Energy Physics (HEP) experiments adopted a distributed computing model more than a decade ago. WLCG, the global computing infrastructure for LHC, in partnership with the US Open Science Grid, has achieved data management at the many-hundred-Petabyte scale, and provides access to the entire community in a manner that is largely transparent to the end users. The main computing challenge of the next decade for the LHC experiments is presented by the HL-LHC program. Other large HEP experiments, such as DUNE and Belle II, have large-scale computing needs and afford opportunities for collaboration on the same timescale. Many of the computing facilities supporting HEP experiments are shared and face common challenges, and the same is true for software libraries and services. The LHC experiments and their WLCG-partners, DUNE and Belle II, are now collaborating to evolve the computing infrastructure and services for their future needs, facilitated by the WLCG organization, OSG, the HEP Software Foundation and development projects such as HEP-CCE, IRIS-HEP and SWIFT-HEP.  In this paper we outline the strategy by which the international HEP computing infrastructure, software and services should evolve through the collaboration of large and smaller scale HEP experiments, while respecting the specific needs of each community. We also highlight how the same infrastructure would be a benefit for other sciences, sharing similar needs with HEP. This proposal is in line with the OSG/WLCG strategy for addressing computing for HL-LHC and is aligned with European and other international strategies in computing for large scale science. The European Strategy for Particle Physics in 2020 agreed to the principles laid out above, in its final report.

## Introduction

High Energy Physics (HEP) experiments started adopting distributed computing models two decades ago. One of the main drivers was the large resource needs of the CERN LHC experiments, together with the funding model and expectations of the contributing countries, at the start of a new digital era. HEP has demonstrated a unique capability with the global computing infrastructure for the LHC, achieving the management of data at the many-hundred-Petabyte scale, and providing access to the entire community in a manner that is largely transparent to the end user. This is still a rather unique facility in science, but as other communities' needs grow beyond what can be provided for by individual facilities, they too started to tackle similar issues. HEP has a challenge for the foreseeable future – which is how to achieve a scale of computing and data management that is one order of magnitude greater than that of today, while maintaining a reasonable cost envelope. The High Luminosity LHC program (HL-LHC) presents, in terms of scale, the largest of these challenges. Other high data-rate experiments, such as Belle II, as well as future facilities, such as the Long Baseline Neutrino Facility with the DUNE experiment must also be considered. The Belle II and DUNE collaborations are building their computing systems leveraging aspects of the infrastructure deployed for LHC. They are customizing it for their needs, contributing to the development of existing services and to new solutions. We believe we should facilitate this process and evolve the current Worldwide LHC Computing Grid (WLCG) infrastructure into a HEP-wide scientific data and computing environment, available for the future to interested parties in our field. Importantly, in addition, we observe similar needs arising in related fields (astronomy, astro-particle) with many of the HEP facilities often directly involved. In planning for the future, we must consider compatibility and synergy at the facility level. Taking the success of the WLCG as a starting point it could be envisaged to evolve the infrastructure and tools as a basis for computing for HEP for the coming years, while addressing the concerns of cost (both in terms of equipment and operational), organization, and community needs.

In parallel with these infrastructure developments, considerable efforts from the experiments and common software projects on a software "upgrade" are underway that are optimizing or rewriting codes for the challenges of the future, to improve their efficiency on modern hardware platforms. In particular, these developments aim to help software run efficiently on heterogeneous hardware, including GPUs, which are now deployed at many WLCG sites and in HPC centers.

Finally, the aspect of sustainability should not be forgotten. Twenty years ago, when Grid computing was invented, there were no examples of how to build such a distributed system, and no experience from industry or others. If we were to design the system today, of course we would benefit from the tools and expertise of the global internet companies. In order to guarantee the sustainability of the infrastructure, a continuous process of modernisation needs to be carried on. New technologies from large open source communities and industry must be regularly evaluated and possibly integrated. Different kinds of facilities such as commercial and academic clouds and HPC centers will play a role in the future computing models, offering a heterogeneous landscape of compute and storage architectures. We need to make sure they can become part of the Grid infrastructure for HEP and we can use them efficiently.

# HEP challenges for the next decade

The demand for scientific computing capacity in HEP continues to increase. For the LHC physics program, the compute core hours consumed by the four experiments as a function of time is shown in Fig. 1. Already, during LHC Run-2 (2015-2018) the funding agencies of the LHC experiments made it clear that funding beyond a constant budget should not be expected during Run-2 and beyond. Such a level of flat funding enabled the provisioning of enough resources for physics at Run-2 and we expect it to be also adequate for Run-3, but it will likely be insufficient for HL-LHC unless a major evolution in the application software, the computing services and the infrastructure happens. The most recent public resource projections for the needs of the ATLAS and CMS experiments at HL-LHC are shown in Fig 1. They are compared with the expected available resources in the flat budget scenario, considering a realistic moderate increase in the computing performance per unit cost over the coming years. Without the evolution mentioned above, and because of the flat funding constraints, WLCG would not be able to cope with the future needs of the LHC experiments.

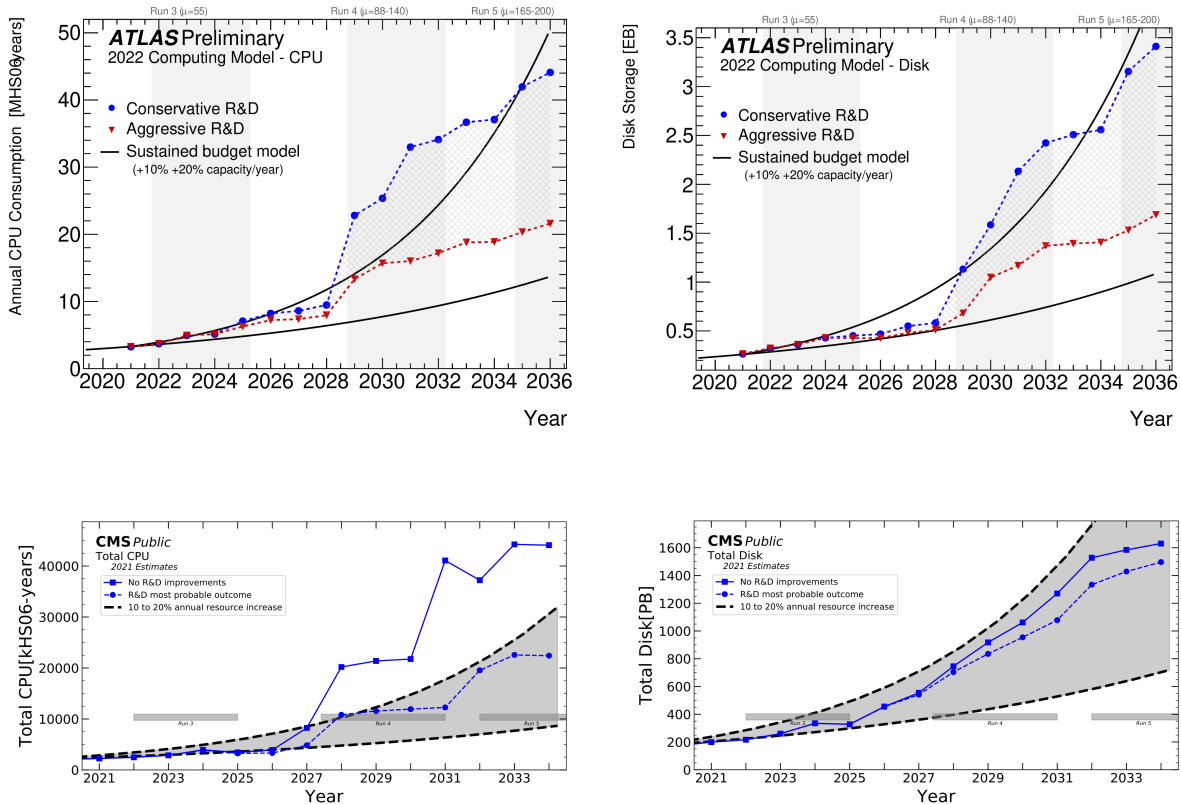

*Figure 1: the ATLAS (top) and CMS (bottom) projections for the computing (left) and disk storage (right) needs for the HL-LHC. The CMS projections were produced in Fall 2021 based on the expected LHC schedule at that time. They present both a scenario with no improvement from R&D and a scenario with the most probable improvements from R&D. The ATLAS projections were produced in early 2022 based on a revisited LHC schedule with a later start of HL-LHC by 18 months. They show projections in a conservative R&D scenario and an aggressive R&D scenario.*

The Belle II experiment started its Physics run in 2019 and aims to collect 50 ab^-1 by the end of data taking in the 2030s, improving on its predecessor by two orders of magnitude. Thanks to its small event size of around 10kB it will produce a fairly modest data volume by HL-LHC standards but will compromise $10^{12}$ real data events and a similar number of simulated events. As the Belle II analysis data format is effectively an augmented version of the reconstructed data format, it presents many of the same challenges as the analysis data formats of CMS and ATLAS.

The DUNE experiment will begin taking data with their first two 10 kT far detector modules in 2028-2029. Each 14x12x58 meter Liquid Argon TPC will produce 3 to 8 GB of raw data per readout with a total expected volume of 30 PB/year, dominated by calibration samples. The collaboration has successfully simulated and reconstructed data from prototype runs at CERN using WLCG infrastructure across 48 international sites. While the CPU needs of the full DUNE experiment will be small on the scale of the LHC experiments, the memory footprint currently necessary for simulation and reconstruction of these very large event data structures challenge the existing WLCG computing resources.  An additional challenge will be reconstruction of supernova candidates, which generate 600 TB of data over 100s. Real supernovae may occur once per century but candidates are expected monthly and, if detected and reconstructed quickly, could provide rough pointing information before the optical signal emerges.

We also note that several projects in the field of Astronomy and Astro-particle physics are expected to start collecting data in the next decade. They share a large fraction of the same computing challenges with HEP and several of the facilities providing storage and compute. Those projects also embrace a distributed computing model and their needs are sometimes comparable to the largest HEP experiments. For example, the Square Kilometer Array (SKA) experiment expects to collect and process exabytes of data, comparably to the HL-LHC experiments and in the same timeframe[1].

# Strategic Directions

Our vision to address the HEP computing challenges for the next decade articulates around three main pillars:

- Strengthen the backbone of core services and policies established in the context of WLCG. They have demonstrated the capability to serve the needs of the HEP community, particularly the LHC experiments, DUNE and Belle II.
- Evolve the WLCG infrastructure to more flexibly integrate modern facilities, such as HPC centers and commercial or academic clouds. Develop and evolve WLCG services to obtain maximum benefit from non-HEP-specific modern standards and solutions from the open source communities. Modernize the offline software libraries to adapt to a rapidly evolving heterogeneous computing architecture landscape.
- Strengthen the collaborative channels between the HEP experiments in the area of software and computing. Promote the adoption of common tools, services, design practices and policies across the experiments. At the same time, retain the flexibility to complement the common services with specific solutions, addressing individual needs. Facilitate the process of evaluating initially specific solutions for more general needs. Foster an inclusive environment where other sciences with similar needs can collaborate and contribute.

Formally, the WLCG is a collaboration of sites providing the infrastructure for the LHC computing needs. The WLCG community is far broader and includes experiments, and middleware and software providers. The next sections will elaborate our collective strategic view.

---

[1] T. An, Science opportunities and challenges associated with SKA big data. Sci. China Phys. Mech. Astron. **62** (2019) 989531.

# Consolidation

The underlying core services of the infrastructure constitute today one of the major values of WLCG[2]. For LHC, computing and storage resources are deployed at close to 200 sites using this infrastructure. The sites providing resources to the LHC experiments, DUNE and Belle II almost entirely overlap. The storage and compute capacity at the sites is provided to the experiments through well established and agreed interfaces and protocols. The compute capacity is generally accessible through the HTCondor and ARC Computing Element. The storage elements offer the HTTP/WebDAV and Xrootd protocols for data management and data access. Xrootd is also the synchronous file streaming service currently in use. The application software is made available through the CVMFS technology, with caching layers deployed at most sites. Most of the experiments rely on the WLCG File Transfer Service (FTS) for bulk and asynchronous data transfer. Such a service allows sharing of network resources and throttling of storage access. To summarize, the same grid middleware stack is used by most HEP experiments with little specific customisation needed.

The baseline grid services are supported by mature monitoring, operational and support processes and teams, including worldwide collaboration on security and incident response. The global Authentication, Authorisation, and Identity management (AAI) service, relies on trust and policy networks. The security incident response coordination provides worldwide expertise in managing and anticipating security threats.

In the US, WLCG relies on the effort from Open Science Grid (OSG) in providing specific middleware and, very importantly, coordinating the deployment and operations of grid services, for HEP and other communities. OSG, for example, is a major partner of the incident response team mentioned above. We believe that OSG plays a very important role in the HEP computing ecosystem and we expect it will continue playing such a key role in its evolution.

WLCG has in place global networking infrastructures, not only those provided by the National Research and Education Networks (NRENs) and their coordinating bodies, but HEP-specific structures such as the LHC Optical Private Network (LHCOPN), and the very successful LHC Open Network Exchange (LHCONE) overlay network, which provide the ease of management and connectivity that will be essential for the future. Today this is already used by more than the LHC experiments, primarily DUNE and Belle II. In the case of LHC network resources, originally expected to be the main limiting factor, they emerged as probably the most solid service, both in terms of capacity growth and reliability. They will play a central role in the evolution of the infrastructure also in future. WLCG has a strong established relationship with the ESNET and Internet2. WLCG participated in the ESNET planning exercise in 2020 and extrapolated from its conclusions the network evolution model[3] for HL-LHC. The strong collaboration between WLCG and the NRENs will be an important enabling factor for the future computing models.

---

[2] S. Campana, I. Bird, and B. Panzer-Steindel, Overview of the WLCG strategy towards HL-LHC computing - April 2020 LHCC review. doi:10.5281/zenodo.5499655 (2021).

[3] S. Campana, WLCG data challenges for HL-LHC - 2021 planning. doi:10.5281/zenodo.5532452 (2021).

To conclude, we have seen LHC and other HEP experiments, such as DUNE and Belle II benefiting from the WLCG infrastructure already. The needs of DUNE and Belle II are in line with the priorities of the WLCG infrastructure and there is therefore a strong motivation for it to be a shared resource, consolidated around its main core services and functionalities. We should note that this paper is not proposing to use the same resources for all experiments, but rather to try and use the same underlying infrastructure, tools, software, and support as far as possible so that new projects are easier to support on existing facilities. Of course, this helps opportunistic use and sharing, but does not impose it.

# Evolution

The inception of the Grid model happened more than twenty years ago. No technology for distributed computing at the required scale was available at that time. WLCG was built leveraging mostly middleware development projects in the US and Europe. That middleware evolved constantly and allowed the LHC experiments to successfully complete the Run-1 and Run-2 physics programs, from 2009 until now. WLCG and partners such as DUNE and Belle-II need, however, to face the challenge of middleware tool evolution and service sustainability. The computing landscape outside HEP evolved considerably in the last decade. Distributed storage and compute is now a standard implementation of many commercial vendors rather than a HEP-specific solution. In addition, many open source communities now develop products that suit the needs of the HEP community and those products are widely adopted and customized at the scale needed by the current and future HEP experiments.

The "grid" that enables the coherent use of resources must evolve over the coming years, and be capable of supporting continually evolving computing models, and be agile to technology changes. WLCG undertakes a program of innovation with the intent to integrate modern solutions from open source communities, while contributing to shaping those solutions for its specific needs. The examples are numerous. The gridFTP protocol, in use in practice only by the HEP experiments for data transfer, is being deprecated and the more standard HTTP protocol will be used from now on. The AAI infrastructure, currently based on X509 certificates, is transitioning to the use of tokens and to the model of federated identities. The US, thanks to the efforts from the experiment operations teams, the WLCG sites and OSG, has been a pioneer of many of those modernizations and we expect this role to be continued and properly funded in future.

The WLCG infrastructure provides a sophisticated way of dealing with all aspects of distributed data management, including the distribution, resilience, archiving and cataloging of the huge volumes of data; and the means to match compute resources to data across globally distributed compute resources. Providing the capability to store, manage, curate and access data is probably the main asset of a HEP computing infrastructure and also one of the main cost drivers. These aspects deserve special attention in defining a strategy for the future. The currently envisaged model[4] builds on the experience of large commercial cloud providers, as well as the LHC expertise in many-hundred-Petabyte scale data management. It is foreseeable on the 2030s timescale to connect most of the large HEP data centers with dedicated and private multi-Tb/s network links.

---

[4] S. Campana, WLCG Strategy towards HL-LHC. doi:10.5281/zenodo.5897018 (2021).

The combined distributed system would store all of an experiment's data, and by policy replicate it between the data centers as needed. In this way, we would achieve reliability and availability of the data as well as ease of management. Connected to this data cloud would be compute resources. These resources may be co-located at the data centers, or may be other facilities, such as HPC facilities, commercial centers or other large-scale, HEP-owned resources. The model also describes the possibility of inclusion of commercially procured storage. Policy and practical reasons would prevent reliance on the latter for non-reproducible data sets, and such storage should be redundant enough that a commercial center could "unplug" without loss of vital data. The data can be processed at the centers hosting it or externally through a content delivery system, minimizing the possible impacts due to network latency or capacity. This architecture clearly relies on a very strong collaboration with the networking community, with adequate policies and capabilities to agilely connect to any commercial partners. This more flexible model would permit the increased use of economical high latency media, such as tapes, as an active store for organized analysis, again helping with cost. This type of model also allows cost optimization through the use of hybrid centers: HEP owning compute resources at a level that is guaranteed to be fully used is very cost effective, and supplementing this with elastically provisioned resources. This would allow an agile control of the cost, and could evolve as the commercial markets evolve. In the future we might move towards a model where sites can provide different agreed capabilities and specialize; this happened in some respects already in specific situations. Depending on the type of resource, some centers may be best suited to specific types of workload and the data cloud mode described above should enable enough flexibility to use them effectively.

Enabling a data cloud (or data lake) model has been one of the directions of the WLCG strategy for HL-LHC[5]. The DOMA[6] activity in WLCG and funded projects such as IRIS-HEP in the US went through a series of R&D activities to define the building blocks of the model and build prototypes. Some of the building blocks are services in use by the HEP experiments in production for a long time, such as FTS and Rucio[7]. Particularly, the Rucio scientific data management software, initially developed for the purposes of the ATLAS experiment, provided a high level of flexibility that allowed it to satisfy the needs of several HEP experiments (including ATLAS, CMS, DUNE and Belle II). It is now in use by those experiments as a high level data organization and orchestration layer. Other solutions such as xCache/StashCache[8] are now becoming part of the service offering in WLCG as caching solutions but also as part of a more sophisticated content delivery layer. Several elements of the data cloud model have been considered and adopted by other sciences. For example the ESCAPE project[9] prototyped a data cloud pilot for the astronomy and HEP sciences in the cluster. While ESCAPE is a EU funded project, its scope spans sciences beyond the European domain, such as HL-LHC and SKA, as well as LIGO as an observer.

The WLCG infrastructure already integrates heterogeneous resources, such as commercial and academic clouds, HPC facilities and volunteer computing. As already mentioned, those facilities will play a more significant role in the future and further harmonization in their adoption is a key element of our strategy. Particularly, HPC centers are expected to play an important role in future HEP computing models. HEP applications are mostly characterized by event-level parallelism that does not require High Performance Computing center's high-speed interconnects. However, HEP could benefit from the economy-of-scale of large HPC centers and its use case is a good fine-grained complement to the large multi-node allocation needs of other sciences. The HPC centers have been so far accessible through interfaces and policies that do not always allow HEP to optimize its use of these resources, in particular for high throughput computing. At the same time, the HEP data access and processing scenarios, characterized by the large volumes of data to ingress and egress are a pattern that HPC centers are preparing for, as it will be more common in future also for other sciences. Other aspects such as the capability to instantiate the workload and the possibly limited external network connectivity also present a challenge for those resources. It is therefore of strategic interest for the HEP experiments, WLCG and the HPC centers to continue collaborating in understanding how infrastructure, services and policies can evolve to optimize the efficient access and use of resources. An evaluation of cost and benefits would need to be done to find the right balance of adoption of HPC resources.

The utilization of resources at HPC facilities comes with challenges not only at the the level of data access, policies and resource scheduling, but also, and more importantly, at the level of application software. More generally, preparing for scientific computing in the next decade, application software will play a critical role. In fact, innovative solutions will need to be considered and implemented: experts with different skill-sets, such as parallel programming and machine learning, will need to complement the more physics-oriented expertise today available in our community.

Improving software performance will be a critical aspect of the way to reduce the cost of computing in the future. The HSF Community White Paper[10] identified key areas in the software domain which should be top priority in the future strategy of HEP computing with particular impact on WLCG on the timescale of HL-LHC (2029). The software frameworks and algorithms of the WLCG experiments were designed many years ago and today cannot efficiently leverage all features and architectures of modern hardware (e.g. vectorization and use of accelerators). Modernizing the software in this direction requires skill sets not broadly available in the HEP community. Building such know-how requires dedicated effort in terms of training knowledgeable domain scientists in software engineering skills. It also needs the right form of recognition in terms of career opportunities for the software developers for which HEP is hardly competitive with industry and therefore has a problem in retaining expertise. It further requires a set of tools and procedures facilitating the software development process, such as elements of the build systems, tools for documentation, and advice on licensing among several others. Existing initiatives such as the HEP Software Foundation (HSF) should be leveraged. Funded projects such as IRIS-HEP and HEP-CCE in the US and SWIFT-HEP in the UK are playing an important role in the modernization of HEP software. HEP must recognise that software efficiency and performance will be key to maintaining an affordable infrastructure. This is not a one-off effort, but will require sufficient and on-going investment in people's skill development and retention. Future funding in the software area will be as important as in the infrastructure to address the future challenges of HEP.

---

[10] The HEP Software Foundation, J. Albrecht et al., A Roadmap for HEP Software and Computing R&D for the 2020s. Comput. Softw. Big Sci. **3** (2019) 7. doi:10.1007/s41781-018-0018-8.

## Collaboration

The WLCG collaboration has demonstrated and implemented a distributed computing model for the LHC experiments, which has played a crucial role in their scientific mission. The WLCG community is strongly established and a network of trust exists between stakeholders. This allows the implementation of a very lightweight decision-making process, based on consensus at various levels of the organization. The non-contractual nature of the agreements WLCG is built upon are in sharp contrast with the commercial computing infrastructures that now exist at similar or much larger scale. The infrastructure, while heterogeneous, builds upon a set of agreed interfaces to compute and storage, common middleware tools and policies in matters of security, identity management, resource sharing, monitoring and accounting.

WLCG has so far been the major player among high energy physics and many other sciences in terms of data volume and compute capacity. This allowed it to steer the evolution of the infrastructure and services in the direction of the LHC experiments' needs. WLCG has been characterized by a pragmatic ability to accommodate the different and changing requirements of the experiments with time and the evolution of their computing systems. In addition, it demonstrated its capability to support applications from very different domains than HEP. An example is the campaign providing resources, both hardware and software, to COVID related studies in the initial phases of the pandemic in 2020.

In the previous section we highlighted how, going forward, other HEP experiments will play a major role in the distributed computing landscape, with Belle II and DUNE as main examples. In order to maximize the return on investment of the Funding Agencies it would be advantageous to foresee, where appropriate, a common infrastructure and set of tools serving the needs of the HEP experiments and, more broadly, sciences they support. A close collaboration between WLCG and other communities is required in order to help ensure the infrastructure is able to effectively meet the requirements of all stakeholders.

Today there is no formal HEP, or more generally, scientific computing collaboration in the sense of governance. The WLCG collaboration has evolved however to create partnership with other HEP experiments and is considering adapting its governance structure to facilitate that partnership. The evolution of the WLCG collaboration, bringing in new communities and new ideas, naturally provides an opportunity to revisit the relationships between stakeholders. The Open Science Grid project in the US is an obvious opportunity to foster collaboration across different sciences. Ambitious initiatives such as the European Open Science Cloud (EOSC) also provide a timely opportunity to rethink how an evolving WLCG can most effectively collaborate with other sciences in HEP and beyond. In general, as we evolve the infrastructure for HEP computing, there is an opportunity to see which parts of the infrastructure address the common needs of the experiments and which would be also of interest to other communities, how the infrastructure can be evolved to better match the needs of others, and thus achieve a more cost-effective overall solution.

One of the major challenges we identified across the HEP experiments is the effort on software development. It is essential that this aspect is recognised and supported. Part of the software work is experiment and community specific. Another part can be achieved through a common effort at different levels: from common tools and procedures, to sharing methodologies, to actual software libraries and toolkits shared by multiple experiments. We proposed in the section above the idea of a common infrastructure for scientific computing, intended as a common set of tools, services and support for experiments to use, with no imposition of particular choices. Much the same can be said of application software, in which a number of very successful projects have provided common experiment independent tools for many years (ROOT, Geant4, Pythia, to name but a few). The HSF has encouraged further growth in this direction, addressing many aspects of software for the HEP community and providing guidance. Its role in reviewing the DUNE future framework requirement document is a clear example[11], as well as fostering common HEP efforts to utilize data science tools[12] and addressing the computational challenges of event generators[13].

Historically software engineering posts have not been given adequate priority, and consequently funding lines have not supported these in a systematic way. This is not tenable for the scale and technical challenges of the future. A major change is needed in the particle physics funding mindset. Adequate support for training initiatives is an important aspect of this: there is a growing consensus that the kind of professionals needed in the field of scientific computing are not just the traditional computer scientists, but rather a profile of researcher with competences crossing the domains of computing as well as physics. This kind of professional, in a small part, already exists in young researchers with a particular attitude towards computing, but their number needs to increase with a new generation, nurtured from the beginning with a specific training in the required disciplines.

## Conclusions

HEP has the opportunity to leverage its strengths and play a central role in the evolution of scientific computing, as it is the community with the largest experience in large scale distributed scientific computing. The HEP computing evolution strategy is based on three pillars: 1. Consolidating the current HEP computing infrastructure; 2. Evolving it to leverage modern technologies and more heterogeneous systems; 3. Fostering the collaboration between HEP experiments and possibly other sciences to consolidate and innovate coherently but flexibly. The strong collaboration between the LHC experiments, DUNE and Belle II on the foundations of the WLCG infrastructure are a starting point, with the plan to be inclusive and not prescriptive with other HEP experiments and sciences.

---

[11] M. Clemencic et al., DUNE Software Framework Requirements - HSF Review. doi:10.5281/zenodo.6323536 (2022).
[12] J. Pivarski et al., HL-LHC Computing Review Stage 2, Common Software Projects: Data Science Tools for Analysis. arXiv:2202.02194 [physics.data-an] (2022). doi:10.48550/arXiv.2202.02194.
[13] The HSF Physics Event Generator WG, A. Valassi et al., Challenges in Monte Carlo Event Generator Software for High-Luminosity LHC. Comput. Softw. Big Sci. 5 (2021) 12.